\documentclass[conference]{IEEEtran}

\usepackage[utf8]{inputenc}
\usepackage{amsmath,amssymb,amsfonts}
\usepackage{algorithmic}
\usepackage{graphicx}
\usepackage{textcomp}
\usepackage{xcolor}
\usepackage{cite}

\begin{document}

\title{AceleradorSNN: A Neuromorphic Cognitive System Integrating Spiking Neural Networks and Dynamic Image Signal Processing on FPGA}

\author{\IEEEauthorblockN{Daniel Gutierrez, Ruben Martinez, Leyre Arnedo, Antonio Cuesta, Soukaina El Hamry}
\IEEEauthorblockA{\textit{Intigia R\&D Department} \\
Alicante, Spain}}

\maketitle

\begin{abstract}
The demand for high-speed, low-latency, and energy-efficient object detection in autonomous systems—such as advanced driver-assistance systems (ADAS), unmanned aerial vehicles (UAVs), and Industry 4.0 robotics—has exposed the limitations of traditional Convolutional Neural Networks (CNNs). To address these challenges, Intigia has developed AceleradorSNN, a third-generation artificial intelligence cognitive system. This architecture integrates a Neuromorphic Processing Unit (NPU) based on Spiking Neural Networks (SNNs) to process asynchronous data from Dynamic Vision Sensors (DVS), alongside a dynamically reconfigurable Cognitive Image Signal Processor (ISP) for RGB cameras. This paper details the hardware-oriented design of both IP cores, the evaluation of surrogate-gradient-trained SNN backbones, and the real-time streaming ISP architecture implemented on Field-Programmable Gate Arrays (FPGA).
\end{abstract}


\section{Introduction}
The evolution of autonomous navigation and industrial inspection requires vision systems capable of operating in real-time under highly variable lighting conditions while maintaining strict power constraints. Conventional Deep Neural Networks (DNNs) and CNNs process continuous frame data, leading to high computational overhead and significant energy consumption. Furthermore, standard RGB sensors struggle with high-speed motion and dynamic range. 

To overcome these limitations, event-based cameras (Dynamic Vision Sensors, DVS) have emerged as a powerful alternative. DVS pixels respond asynchronously to changes in illumination, offering microsecond latency and high dynamic range. However, event data lacks the static, high-resolution texture and color information critical for certain classification tasks. 

The AceleradorSNN project introduces a synergistic architecture that merges the advantages of DVS and RGB sensors. By employing a Neuromorphic Processing Unit (NPU) utilizing Spiking Neural Networks (SNNs) to process DVS events, the system achieves ultra-fast object detection. Crucially, this NPU acts as a cognitive controller, generating real-time adjustment instructions for a Cognitive Image Signal Processor (ISP), which dynamically reconfigures the RGB camera parameters to adapt to environmental changes. Designed for FPGA and ASIC implementation, the system achieves a high Technology Readiness Level (TRL 6/7) tailored for critical embedded applications.

\section{Related Work}
The integration of event-based vision and neuromorphic computing has been extensively explored in recent literature. Gallego et al. highlighted the advantages of event cameras for high-speed tracking, though noting the challenges of their sparse, asynchronous nature \cite{gallego2020}. To process this data, SNNs offer a biologically inspired approach, communicating via discrete spikes rather than continuous values, significantly reducing energy consumption. 

Cordone et al. demonstrated the viability of adapting traditional CNN architectures (such as VGG and DenseNet) into the spiking domain for automotive event data \cite{cordone2022}. Fan et al. further advanced this with the Spiking Fusion Object Detector (SFOD), proving that SNNs can achieve robust multiscale object detection \cite{fan2024}. 

In parallel, hardware-efficient ISP designs are critical for embedded vision. Algorithms like the Malvar-He-Cutler linear image demosaicing \cite{getreuer2011} and Non-Local Means (NLM) denoising \cite{feng2021}, specifically adapted for FPGA architectures by Koizumi and Maruyama \cite{koizumi2020}, provide the foundation for low-latency spatial image processing without relying on external memory buffers. Additionally, dynamic defective pixel correction techniques, such as those proposed by Yongji and Xiaojun \cite{yongji2020}, are essential for maintaining sensor fidelity in harsh environments.

\section{System Architecture Overview}
The AceleradorSNN framework is divided into two primary Intellectual Property (IP) cores, designed to operate synchronously on an FPGA platform:

\begin{enumerate}
    \item \textbf{Neuromorphic Processor (PNN):} The central AI component. It implements an SNN to process DVS events in real-time, detecting dynamic objects and generating parameter adjustment instructions based on the scene's lighting and motion profile.
    \item \textbf{Cognitive ISP:} A fully pipelined hardware module that receives the control instructions from the PNN and dynamically adjusts its internal algorithms (e.g., exposure, white balance) to optimize the high-resolution output of a standard RGB sensor.
\end{enumerate}

By combining these modules, the system achieves maximum performance in perception precision, versatility, and adaptability to changing conditions, all while minimizing energy consumption.

\section{Neuromorphic Processing Unit (NPU) Design}
The NPU is engineered to extract spatial and temporal features from DVS data using spiking neurons.

\subsection{Event Encoding}
Raw DVS events are represented as tuples containing the timestamp, spatial coordinates, and polarity: $e = (t, x, y, p)$. Because SNNs operate in discrete time steps, the continuous asynchronous stream is segmented into fixed temporal windows \cite{detournemire2020}. Within each window, events are aggregated into temporal bins and encoded using a one-hot spatial-temporal voxel grid. This transformation yields a multidimensional tensor representing time steps, polarity channels, and spatial dimensions, which serves as the input to the spiking convolutional layers.

\subsection{Neuron Model and Training}
The architecture utilizes the Leaky Integrate-and-Fire (LIF) neuron model, balancing biological realism with computational efficiency. The LIF neuron simulates the natural loss of charge via a decay term. Its membrane potential $u(t)$ is governed by the differential equation:

\begin{equation}
\tau_m \frac{du(t)}{dt} = u_{rest} - u(t) + RI(t)
\end{equation}

where $\tau_m$ is the membrane time constant, $R$ is resistance, and $I(t)$ is the input current. A spike is emitted when the membrane potential reaches a defined threshold, after which the potential is reset. Because the discrete firing event is non-differentiable, training is conducted using Surrogate Gradients, which approximate the derivative of the spike function. This allows the use of Backpropagation Through Time (BPTT) and standard optimizers like AdamW to update network weights effectively.

\subsection{Backbone Evaluation}
To optimize the NPU, three adapted spiking architectures were evaluated using the Prophesee’s GEN1 Automotive Detection Dataset (Prophesee GEN1):
\begin{itemize}
    \item \textbf{Spiking-VGG:} A deep, uniform sequence of convolutional layers ideal for hierarchical feature extraction.
    \item \textbf{Spiking-DenseNet:} Features dense blocks where the output of each layer feeds into all subsequent layers, preventing gradient vanishing and promoting feature reuse.
    \item \textbf{Spiking-MobileNet:} Utilizes depthwise separable convolutions to drastically reduce parameter count and computational cost.
    \item \textbf{Spiking YOLO:} Converts the standard YOLO architecture into a Spiking Neural Network (SNN), utilizing temporal spikes rather than continuous values to achieve high performance on neuromorphic hardware.
\end{itemize}

Experimental results with quantized models indicated that Spiking YOLO achieved the highest overall precision (Average Precision of 0.4726 at IoU 0.50), presenting the best balance between accuracy and computational cost. Conversely, Spiking-MobileNet exhibited the highest network sparsity (48.08\%), meaning a large proportion of neurons remained inactive, which is highly desirable for maximizing energy efficiency in low-power edge devices.

\section{Cognitive Image Signal Processor (ISP) Design}
The Cognitive ISP is designed entirely in HDL as a modular, parameterizable pipeline. It operates on a continuous data stream, processing pixels individually as they traverse the pipeline without the need to store full image frames, thereby drastically reducing hardware area.

\subsection{AXI4-Stream Integration}
Data transfer between ISP modules relies on the AXI4-Stream protocol, a point-to-point standard ideal for embedded systems. Handshaking is governed by \texttt{tvalid} (master indicating valid data) and \texttt{tready} (slave indicating readiness), ensuring seamless data flow and pipeline stalling when necessary.

\subsection{Pipeline Stages and Algorithms}
The ISP implements a sequence of real-time correction and enhancement stages:
\begin{enumerate}
    \item \textbf{Dynamic Defective Pixel Correction (DPC):} Based on the algorithm by Yongji and Xiaojun \cite{yongji2020}, this module identifies dead pixels by analyzing a $5 \times 5$ spatial window. Line buffers are utilized to cache incoming rows. A pixel is marked defective if its intensity significantly deviates from its neighbors across multiple directional gradients.
    \item \textbf{Auto White Balance (AWB) \& White Balance (WB):} A state machine analyzes the image to calculate RGB gains, discarding overexposed and underexposed pixels before applying the corrective gains dynamically based on NPU feedback.
    \item \textbf{Demosaicing:} Converts Bayer pattern data to full RGB. The hardware implementation utilizes the Malvar-He-Cutler linear interpolation method \cite{getreuer2011}, which estimates missing color channels based on surrounding gradients to preserve sharp edges.
    \item \textbf{Non-Local Means (NLM) Denoising:} To remove Gaussian noise without blurring edges, the system implements an FPGA-adapted NLM algorithm \cite{koizumi2020}. Unlike local filters, NLM relies on structural redundancy, computing the Euclidean distance between a reference patch and neighboring patches within a search window.
    \item \textbf{Gamma Correction \& Color Space Conversion:} Custom LUTs (Look-Up Tables) apply non-linear gamma curves, followed by a configurable fixed-point arithmetic module to convert the RGB signal to the YCbCr color space for independent luminance sharpening.
\end{enumerate}

\section{System Integration and Hardware Deployment}
The complete AceleradorSNN system is unified through a top-level integration module. The architecture allows the NPU and ISP to operate in a closed cognitive loop. The NPU interfaces directly with the DVS sensor logic and processes event tensors via the spiking neural network. Once the NPU detects an object and identifies localized lighting anomalies, it transmits configuration parameters via a control interface to the ISP. 

The ISP's synchronization controller aligns the DVS and RGB data streams. It interprets the NPU's parameter updates—such as modifying the AWB gains, tweaking the Gamma LUTs, or adjusting the NLM denoising strength—applying these changes on-the-fly to the RGB feed. This allows the extraction of high-resolution, context-rich images of the detected objects, overcoming the traditional trade-offs between speed, dynamic range, and image fidelity. 

The entire framework has been synthesized and validated on FPGA, with tools mapping the design for future ASIC fabrication.

\section{Conclusion}
Intigia's AceleradorSNN establishes a new paradigm in embedded artificial intelligence for aerospace, automotive, and industrial applications. By successfully marrying the ultra-low latency and energy efficiency of event-driven Spiking Neural Networks with the high-resolution output of a dynamically controlled Cognitive ISP, the system transcends the limitations of conventional CNNs. The rigorous FPGA implementation of local streaming algorithms and surrogate-gradient trained LIF networks demonstrates a viable, high-TRL pathway toward fully autonomous, highly adaptable, and environmentally sustainable AI edge perception.

\section*{Acknowledgment}
This work has been supported by the program ``Proyectos de desarrollo experimental en el área de la inteligencia artificial y para el impulso al desarrollo de espacios de datos sectoriales'', within the framework of the Recovery, Transformation and Resilience Plan, funded by the European Union - NextGenerationEU, under the project reference number INREIA/2024/62.
This specific grant (INREIA/2024/62) was officially awarded to Intigia S.L. under this NextGenerationEU funding framework to support the experimental development of Artificial Intelligence

\end{document}